\documentclass[usenatbib,twocolumn,a4paper]{mnras}
\usepackage{graphicx}
\usepackage{times}
\usepackage{amssymb}
\usepackage{natbib}
\usepackage{rotating}
\usepackage{lscape}
\usepackage{multirow}
\usepackage{ulem}
\usepackage{bm} 
\usepackage{epsfig}
\usepackage{times}
\usepackage{amsmath}
\usepackage{color}

\title[Non-Linear Matter Power Spectrum Covariance Matrix Errors]{Non-Linear Matter Power Spectrum Covariance Matrix Errors and Cosmological Parameter Uncertainties}
\author[Blot, Corasaniti, Amendola \& Kitching]{L. Blot$^{1,2}$\thanks{linda.blot@obspm.fr}, P.S. Corasaniti$^1$, L. Amendola$^{3}$, T.D. Kitching$^{4}$\\ 
$^1$Laboratoire Univers et Th\'eories, UMR 8102 CNRS, Observatoire de Paris, Universit\'e Paris Diderot, 5 Place Jules Janssen, 92190 Meudon, France\\
$^2$Institut de Ci\`encies de l'Espai, IEEC-CSIC, Campus UAB, c/ Can Magrans s/n, 08193 Cerdanyola del Vall\`es, Barcelona, Spain\\
$^3$Institut f\"{u}r Theoretische Physik, Ruprecht-Karls-Universit\"{a}t Heidelberg, Philosophenweg 16, 69120 Heidelberg, Germany\\
$^4$Mullard Space Science Laboratory, University College London, Holmbury St Mary, Dorking, Surrey RH5 6NT, United Kingdom}

\begin{document}

\maketitle
\begin{abstract}
The covariance of the matter power spectrum is a key element of the analysis of galaxy clustering data. Independent realisations of observational measurements can be used to sample the covariance, nevertheless statistical sampling errors will propagate into the cosmological parameter inference potentially limiting the capabilities of the upcoming generation of galaxy surveys. The impact of these errors as function of the number of realisations has been previously evaluated for Gaussian distributed data. However, non-linearities in the late time clustering of matter cause departures from Gaussian statistics. Here, we address the impact of non-Gaussian errors on the sample covariance and precision matrix errors using a large ensemble of N-body simulations. In the range of modes where finite volume effects are negligible ($0.1\lesssim k\,[h\,{\rm Mpc^{-1}}]\lesssim 1.2$) we find deviations of the variance of the sample covariance with respect to Gaussian predictions above $\sim 10\%$  at $k>0.3\,h\,{\rm Mpc^{-1}}$. Over the entire range these reduce to about $\sim 5\%$ for the precision matrix. Finally, we perform a Fisher analysis to estimate the effect of covariance errors on the cosmological parameter constraints. In particular, assuming Euclid-like survey characteristics we find that a number of independent realisations larger than $5000$ is necessary to reduce the contribution of sampling errors to the cosmological parameter uncertainties at sub-percent level. We also show that restricting the analysis to large scales $k\lesssim0.2\,h\,{\rm Mpc^{-1}}$ results in a considerable loss in constraining power, while using the linear covariance to include smaller scales leads to an underestimation of the errors on the cosmological parameters.
\end{abstract}

\section{Introduction}\label{intro}
The next generation of galaxy surveys will map the distribution of matter in the universe with unprecedented accuracy over large cosmic volumes. Surveys such as the Large Synoptic Survey Telescope\footnote{www.lsst.org} (LSST) and the Euclid mission\footnote{www.euclid-ec.org} are designed to detect millions of galaxies over a wide range of scales and redshifts, potentially providing measurements of the clustering of matter to few per-cent statistical uncertainty. In order to infer unbiased cosmological parameter constraints the analysis of these data will require theoretical model predictions that account for the non-linearities of the late time gravitational collapse of matter. As an example, on the scales of the Baryon Acoustic Oscillations (BAO) deviations from linear predictions are at a few per-cent level \citep[see e.g.][]{Rasera2014}. However, even the availability of such accurate predictions may not suffice to correctly analyse the data, since non-linearities induce mode-couplings which cause errors on band powers to become increasingly correlated \citep{Meiksin1999,Scoccimarro1999}. Because of this, future galaxy survey measurements will require accurate estimation of covariance matrices. 

In the case of the matter power spectrum the covariance matrix can be estimated from large ensembles of N-body simulations \citep{Takahashi2009,Harnois2012,Blot2015}. However, due to finite sampling, the estimation of the covariance is affected by statistical errors that propagates into the cosmological parameter uncertainties. The impact of such errors has been evaluated in a number of studies assuming Gaussian distributed data \citep[see e.g.][]{Taylor2013,Dodelson2013,Percival2014,Taylor2014}. However, as shown by \cite{Blot2015} deviations from the linear clustering regime induce increasingly larger non-Gaussian errors already at the BAO scale. In principle the use of analytical models of the covariance avoids sampling errors \citep[see e.g.][for work in this direction]{TakadaHu2013,Irshad2014}. Nonetheless, such models still need to match numerical simulation results and as shown by \citet[][]{WuHuterer2013} in the context of the halo model, model calibration uncertainties will propagate in the estimation of parameter errors.

In this work we set to assess what is the impact of non-linearities on the statistical errors associated with the sample covariance estimator and what are the effects on the cosmological parameter uncertainties expected from future galaxy surveys. To this purpose we use a large ensemble of N-body simulations to compare covariance matrix estimation errors against model predictions for Gaussian distributed data. Then, assuming Euclid-like survey characteristics, we perform a Fisher matrix analysis to forecast the impact of non-Gaussian errors on cosmological parameter constraints. 

The paper is organised as follows: in Section~\ref{sec1} we introduce the N-body simulation ensemble used in the analysis; in Section~\ref{sec2} we discuss Gaussian prediction of the sample covariance and precision matrix errors and present the comparison with estimates from the N-body simulation ensemble; in Section~\ref{sec_fisher} we present the results of the Fisher analysis and finally we discuss the conclusion in Section~\ref{conclu}.

\section{N-body Simulation Ensemble}\label{sec1}
We consider the simulation ensemble Set A from the Dark Energy Universe Simulation - Parallel Universe Runs (DEUS-PUR) project presented in \citet{Blot2015}. This set consists of $12288$ Adaptive Mesh Refinement (AMR) N-body simulations of $(656.25\,h^{-1}\,\textrm{Mpc})^3$ volume with $256^3$ particles (corresponding to a mass resolution of $m_p=1.2\times 10^{12}$ M$_\odot$ h$^{-1}$ at coarse level) of a flat $\Lambda$-Cold Dark Matter ($\Lambda$CDM) best-fit model to the WMAP-7 years data \citep{Spergel2007}. For each simulation the matter power spectrum is computed in the range $k_{\rm min}\approx 0.01$ $h$ Mpc$^{-1}$ (corresponding to the fundamental mode of the simulation box) to $k_{\rm max}\approx 1.22$ $h$ Mpc$^{-1}$ (half the Nyquist frequency of the grid used to compute the spectra). The spectra have been corrected for mass resolution errors as described in \citet{Blot2015} using simulation Set B consisting of 96 N-body simulations of identical volume with $1024^3$ particles corresponding to a mass resolution of $m_p=2\times 10^{10}$ M$_\odot$ h$^{-1}$ at coarse level.

\begin{figure}
\begin{centering}
\includegraphics[scale=0.27]{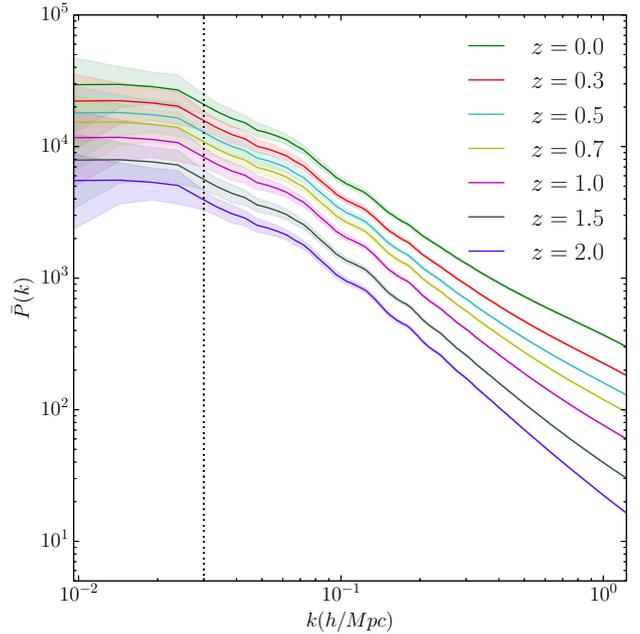}
\caption{Average and standard deviation of the non-linear matter power spectrum at $2\ge z\ge0$ (bottom to top) from DEUS-PUR simulations Set A corrected for mass resolution errors using Set B as described in \citet{Blot2015}. The vertical dotted line indicates the point where cosmic variance errors reduce to less than $10\%$.}
\label{fig1}
\end{centering}
\end{figure}

In Fig.~\ref{fig1} we plot the average matter power spectrum and the standard deviation at several redshifts in the interval $0\le z\le 2$. The standard deviation is largest near $k_{\rm min}$ due to cosmic variance and reduces to less than $10\%$ for $k\gtrsim 0.03$ $h$ Mpc$^{-1}$. Thus, to be conservative from now on we set $k_{\rm min}=0.03$ $h$ Mpc$^{-1}$ unless otherwise specified. 

Using this ensemble of spectra we compute the covariance matrix at each redshift with the sample covariance estimator:
\begin{equation}\label{cov_def}
\widehat{\mathcal{C}}_{ij} = \frac{1}{N_s-1}\sum_{n=1}^{N_s}[\hat{P}_n(k_i)-\bar{P}(k_i)][\hat{P}_n(k_j)-\bar{P}(k_j)],
\end{equation}
where $N_s$ is the number of independent realisations and $\bar{P}(k)=\sum_{n=1}^{N_s}\hat{P}_n(k)/N_s$ is the sample mean, with $\hat{P}_n(k)$ the matter power spectrum of the $n$-th realisation. This is an unbiased estimator of the covariance but, as we later explain, its inverse is not an unbiased estimator of the inverse of the covariance matrix, also called the precision matrix.

\section{Covariance and Precision Matrix Errors}\label{sec2}
Let us consider a set of data consisting of measurements of the matter power spectrum, $P_{k_i}^d$ collected in $i=1,..N_d$ bands. A standard likelihood analysis will use these measurements to infer constraints on a set of parameters $\vec{\theta}$ given a model prediction of the power spectrum $P_{k_i}^t(\vec{\theta})$. Assuming that the density contrast in Fourier space is Gaussian distributed, the power spectrum in each bin of width $\Delta k$ follows a $\chi^2$ distribution with a number of degrees of freedom given by the number of $k$-modes contained in the bin, i.e. $N_k\approx k^2 \Delta k \,V/(2 \pi^2)$, where $V$ is the volume. In linear regime and in the limit of large $N_k$ the power spectrum distribution tends to a Gaussian distribution. In non-linear regime deviations from the $\chi^2$ distribution arise \citep{Blot2015} that may bias the inferred parameter values. Here we concentrate on the effect of non-linearities on the covariance estimator and we leave the study of the impact of the shape of the likelihood on parameter inference for future work.

In the case of Gaussian distributed data the likelihood reads as
\begin{equation}\label{likelihood}
\mathcal{L}\propto \exp{\left\{-\frac{1}{2}\sum_{i,j=1}^{N_d}\left[P_{k_i}^d-P_{k_i}^t(\vec{\theta})\right]\mathcal{C}^{-1}_{ij}\left[P_{k_j}^d-P_{k_i}^t(\vec{\theta})\right]\right\}},
\end{equation}
where $\mathcal{C}^{-1}_{ij}$ is the precision matrix. Notice that in writing Eq.~(\ref{likelihood}) we have assumed that the covariance is independent of the parameters $\vec{\theta}$, in such a case we can neglect the normalisation factor which does not play any role in the determination of the parameter constraints. Nevertheless, it is worth remarking that the power spectrum covariance matrix may indeed vary with the cosmological parameters \citep[see e.g.][who have studied the impact of a cosmological model dependent covariance obtained from simulations of a lognormal galaxy field]{Labatie2012}. However, the extent to which the cosmological parameter inference is affected by cosmological model dependencies of the covariance due to non-linearities is still not known and will require a dedicated study which is beyond the scope of this work. 

The point that we want to address here is how non-linearities of the density field impact the estimation errors of the sample covariance and precision matrix. These errors arise from the fact that the true covariance $\mathcal{C}$ is unknown, we only have an unbiased estimate of it as given by Eq.~(\ref{cov_def}). This has the consequence that the inverse of the sample covariance, $\widehat{\mathcal{P}}\equiv\widehat{\mathcal{C}}^{-1}$, is not an unbiased estimator of the precision matrix due to the noise in the sample covariance \citep{Anderson2003}. Moreover the noise in the sample covariance propagates in the cosmological parameter inference via Eq.\eqref{likelihood}. 

\subsection{Precision Matrix Bias}
Let us first consider the problem of the bias of the precision matrix estimator. This can be built by inverting the sample covariance estimator such that $\widehat{\mathcal{P}}\equiv\widehat{\mathcal{C}}^{-1}$ provided that $N_s>N_d+1$ otherwise the sample covariance is not full rank and the inverse is undefined \citep{Hartlap2007}. For Gaussian distributed data the expectation value of the inverse of the sampled covariance matrix, assuming that the mean of the precision matrix distribution is unknown, is given by \citep[see e.g.][]{Press1982,Anderson2003,Taylor2013}:
\begin{equation}\label{biased_precision}
\langle \widehat{\mathcal{P}}_{ij}\rangle=\frac{N_s-1}{N_s-N_d-2}\,\mathcal{P}_{ij},
\end{equation}
where $N_s$ is the number of simulations used to estimate the sample covariance. It follows that an unbiased estimate of the precision matrix is given by:
\begin{equation}\label{unbiased_precision}
\widehat{\mathcal{P}}^{\rm unbiased}_{ij}=\frac{N_s-N_d-2}{N_s-1}\,\widehat{\mathcal{P}}_{ij},
\end{equation}
which is the only unbiased estimator of the precision matrix \citep{Taylor2013}. This estimator is defined for $N_s>N_d+2$.
We test the validity of Eq.~(\ref{biased_precision}) using the ensemble of $N_t=12288$ spectra from the DEUS-PUR Set A corrected for mass resolution errors and sampled over $N_d=250$ bands in the range $0.03\lesssim k\,[h\,{\rm Mpc^{-1}}]\lesssim 1.22$ at $z=0$. 

\begin{figure}
\begin{centering}
\includegraphics[scale=0.48]{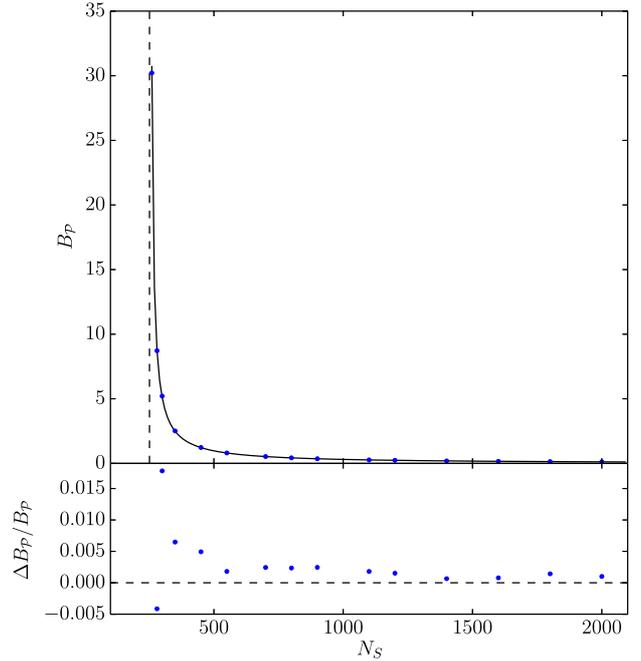}
\caption{Fractional bias of the trace of the mean of the sample precision matrix at $z=0$ as function of the number of simulations, $N_s$. The solid black line is the predicted scaling for Gaussian distributed data, Eq.~(\ref{trace_bias}), while the dots are the estimates from the N-body simulations. The vertical dashed line indicates the minimum number of simulations for which the sample covariance is positive definite. In the bottom panel is shown the relative difference.}
\label{fig2}
\end{centering}
\end{figure}

To compute the average of the sample precision matrix which appears in the left-hand-side of Eq.~(\ref{biased_precision}) as function of the number of simulation $N_s$ we divide the ensemble of spectra in $N_g=\textrm{int}(N_t/N_s)$ groups, where $\textrm{int}(N_t/N_s)$ indicates the quotient of $N_t$ and $N_s$, in each group we estimate the sample covariance using Eq.~(\ref{cov_def}) and computing the inverse we obtain the biased estimate of precision matrix. Then, we compute the average biased precision matrix over the $N_g$ groups as $\langle\widehat{\mathcal{P}}\rangle=1/N_g \sum_k\widehat{\mathcal{P}}_k$ which we compare to that obtained using the entire ensemble of $N_t$ spectra, $\mathcal{P}$, on the right-hand-side of Eq.~(\ref{biased_precision}). To compare the numerical results against the theoretical prediction we follow \citet{Taylor2013} and compute the fractional bias defined as:
\begin{equation}\label{trace_bias}
B_{\mathcal{P}}\equiv\frac{\mathrm{Tr}\,\langle\widehat{\mathcal{P}}\rangle-\mathrm{Tr}\,\mathcal{P}}{\mathrm{Tr}\,\mathcal{P}}=\frac{N_s-1}{N_s-N_d-2}-1,
\end{equation}
which we plot in Fig.~\ref{fig2} as function of $N_s$ from the ensemble of spectra at $z=0$. In the upper panel the solid black line is the scaling expected for Gaussian distributed data, while the blue points are the numerical results, the relative difference is shown in the lower panel. We can see that for $N_s>500$ the analytical prediction agrees to the N-body simulation results to better than $0.5\%$. 

\subsection{Variance of Sample Covariance}
\begin{figure}
\begin{centering}
\includegraphics[scale=0.5]{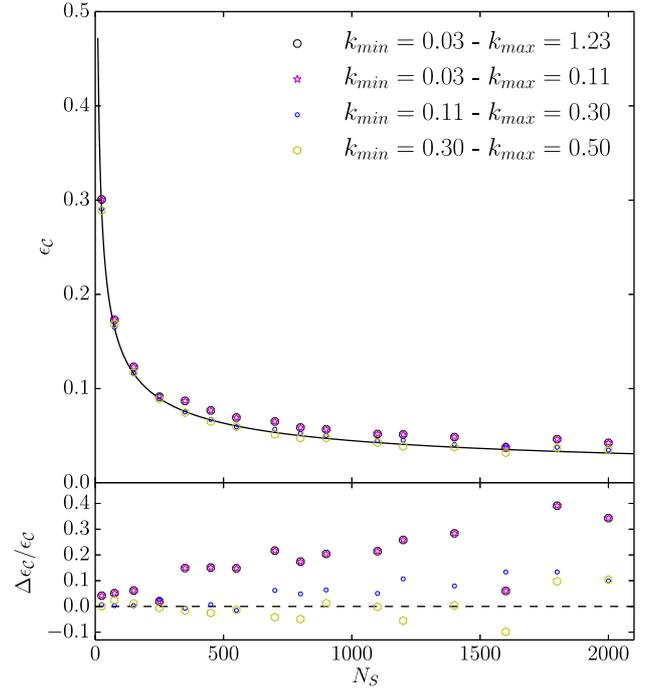}
\caption{Relative sample covariance error $\epsilon_{\mathcal{C}}$ as function of $N_s$ in different wavenumber intervals with increasing values of $k_\textrm{min}=0.03$ (magenta stars), $0.11$ (blue dots) and $0.30$ $h$ Mpc$^{-1}$ (yellow hexagons). Black solid line is the expected scaling from the Wishart distribution. In the bottom panel is shown the relative difference with respect to the expected scaling. The error on the whole diagonal (black octagones) is dominated by the finite volume errors in the interval $0.03<k\,[h\,{\rm Mpc^{-1}}]<0.11$.}
\label{fig3}
\end{centering}
\end{figure}

\begin{figure}
\begin{centering}
\includegraphics[scale=0.5]{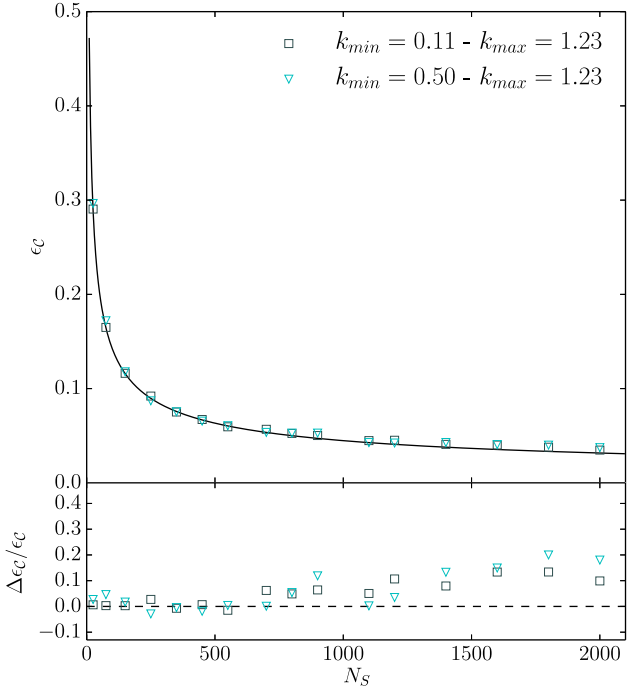}
\caption{As in Fig.~\ref{fig3} for larger $k$ values with $k_\textrm{min}=0.50$ (light blue triangles). Black solid line is the expected scaling from the Wishart distribution. In the bottom panel is shown the relative difference with respect to the expected scaling. When removing the modes below $k=0.11\,h\,{\rm Mpc^{-1}}$ the error on the whole diagonal (grey squares) deviates from the prediction at $\sim10\%$ level due to non-linearities.}
\label{fig4}
\end{centering}
\end{figure}

Let us now consider the errors on the sample covariance. For Gaussian distributed data the statistics of the sample covariance is described by the Wishart distribution \citep{Wishart1928,Press1982}. \citet{Taylor2013} have derived an analytical expression for the variance which reads as:
\begin{equation}\label{variance_of_covariance}
\sigma^2(\widehat{\mathcal{C}}_{ij})=\frac{1}{N_s-1}\left(\mathcal{C}_{ij}^2+\mathcal{C}_{ii}\mathcal{C}_{jj}\right).
\end{equation}
As for the estimation of the fractional bias we estimate the average and standard deviation of the covariance over $N_g=\textrm{int}(N_t/N_s)$ groups of simulations as function of $N_s$. We can test the validity of Eq.~(\ref{variance_of_covariance}) along the diagonal elements by computing the ratio:
\begin{equation}\label{err_cov_scaling}
\epsilon_{\mathcal{C}}=\sqrt{\frac{\sum_i\sigma^2(\widehat{\mathcal{C}}_{ii})}{\sum_i\langle\widehat{\mathcal{C}}_{ii}\rangle^2}}=\sqrt{\frac{2}{N_s-1}},
\end{equation}
which we plot in Fig.~\ref{fig3} as function of $N_s$ having taken the sum over the diagonal elements for three different $k$-intervals. We can see that deviations from the expected scaling in Eq.~(\ref{err_cov_scaling}) are largest ($\gtrsim20\%$) for $0.03<k\,[h\,{\rm Mpc^{-1}}]<0.11$ and decrease for increasing values of $k_\textrm{min}$. These deviations are due to finite volume effects which as shown in Fig.~\ref{fig1} manifest in a larger standard deviation of the estimated power spectra at low $k$. If we sum over the entire diagonal the cumulative error is dominated by these effects, as can be seen by comparing black octagons and magenta stars in Fig.~\ref{fig3}. For $k>0.11$ $h$ Mpc$^{-1}$ this sample variance effect is reduced to less than $5\%$ on the matter power spectrum and correlates with the reduced discrepancy from the scaling of Eq.\eqref{err_cov_scaling} (blue dots). Nonetheless, we may still notice deviations up to $\sim 10\%$ in the higher $k$-interval (yellow hexagons). This can be seen more clearly in Fig.~\ref{fig4} where we consider larger wavenumbers. In particular, we may notice increasing departures from Eq.~(\ref{err_cov_scaling}) above $\sim 10\%$ level for $N_s>1500$. We interpret this systematic trend as an indication of deviations from the Wishart distribution due to the non-linearities of the matter density field which cause non-Gaussian errors. If we exclude the bins dominated by sample variance and we sum over the remaining diagonal elements we can see that the cumulative error is dominated by the errors on non-linear scales, as shown in Fig.~\ref{fig4}.

\begin{figure}
\begin{centering}
\includegraphics[scale=0.5]{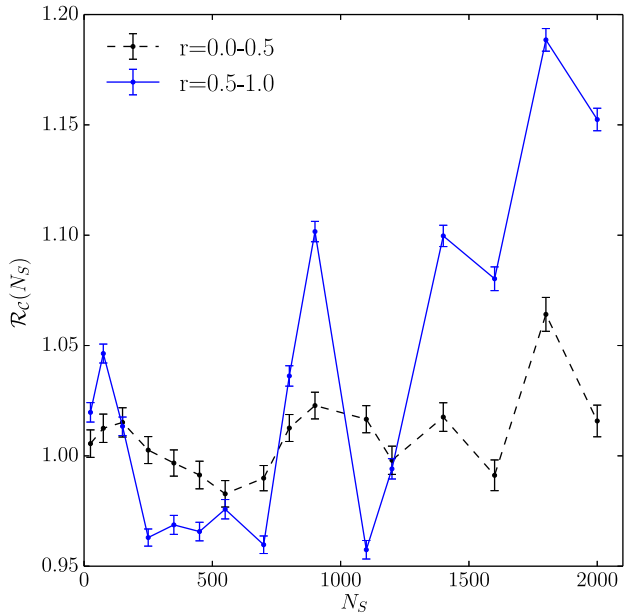}
\caption{Ratio of the sum of the left-hand side of the off-diagonal components of Eq.~(\ref{variance_of_covariance}) relative to that of the right-hand side for wavenumber configurations corresponding to correlation matrix elements with values in the range $0<r<0.5$ (black points) and $0.5<r<1$ (blue points). Error-bars indicate the dispersion.}
\label{fig5}
\end{centering}
\end{figure}

The same trends can be seen in the off-diagonal components of the variance of the sample covariance. To this purpose we estimate the ratio of the sum of the left and right-hand side of Eq.~(\ref{variance_of_covariance}):
\begin{equation}
\mathcal{R}_{\mathcal{C}}=\frac{(N_s-1)\sum_{ij}\sum_{m=1}^{N_g}\left(\widehat{\mathcal{C}}_{ij,m}-\langle\widehat{\mathcal{C}}_{ij}\rangle\right)^2}{(N_g-1)\sum_{ij}\left(\langle\widehat{\mathcal{C}}_{ij}\rangle^2+\langle\widehat{\mathcal{C}}_{ii}\rangle\langle\widehat{\mathcal{C}}_{jj}\rangle\right)}
\end{equation}
over off-diagonal elements for which the corresponding elements of the matter power spectrum correlation matrix:
\begin{equation}
r_{ij} = \frac{\widehat{\mathcal{C}}_{ij}}{\sqrt{\widehat{\mathcal{C}}_{ii}\, \widehat{\mathcal{C}}_{jj}}},
\end{equation}
are below and above the $50\%$ level. We plot the results in Fig.~\ref{fig5}. In the case of off-diagonal elements with correlation $<0.5$ we can see that the ratio is of order unity, while for off-diagonal elements characterised by larger correlations ($>0.5$) the ratio deviates from unity by more than $10\%$ for $N_s>1500$. In the latter case the scatter at low $N_s$ values is large due to large sampling error in the elements of the covariance, but for $N_s>1500$ the deviation exceeds the level of scatter. This clearly shows that on scales where $r>0.5$ the non-Gaussian errors due to the non-linearities of the late-time clustering of matter cause deviations of the sample covariance errors from expectations of the Wishart distribution. It is worth noticing that even with large ensemble of simulations available to us, from the numerical analysis we are unable to asses whether these deviations saturate for very large $N_s$ values, suggesting a significant departure from the Wishart distribution.

\subsection{Variance of Sample Precision Matrix}
\begin{figure}
\begin{centering}
\includegraphics[scale=0.5]{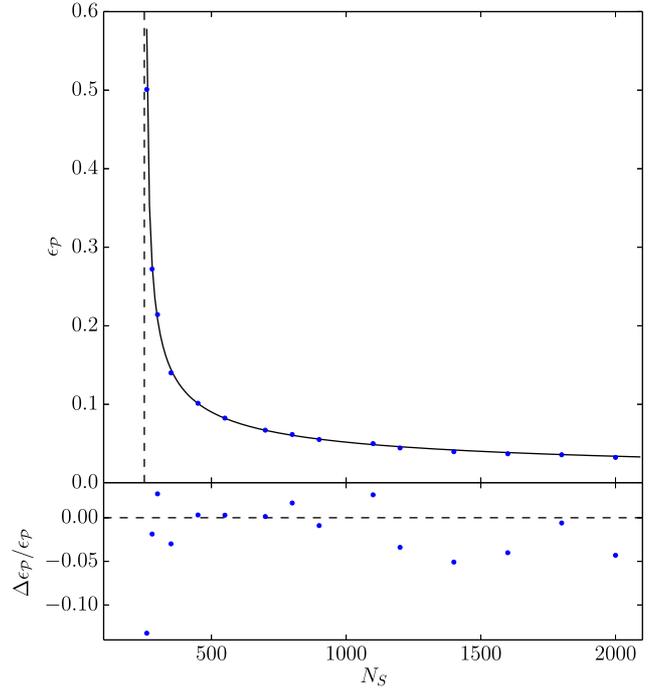}
\caption{Relative sample precision error $\epsilon_{\mathcal{P}}$ as function of $N_s$ in the range $0.11<k\,[h\,{\rm Mpc^{-1}}]<1.22$. Black solid line is the expected scaling from the inverse-Wishart distribution. The vertical dashed line indicates the minimum number of simulations for which the sample covariance is positive definite. In the bottom panel is shown the relative difference with respect to the expected scaling.}
\label{fig6}
\end{centering}
\end{figure}

\begin{figure}
\begin{centering}
\includegraphics[scale=0.5]{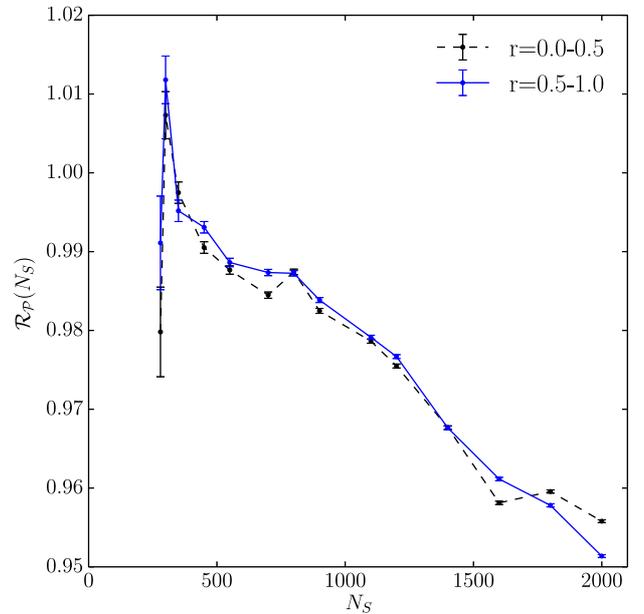}
\caption{As in Fig.~\ref{fig5} for the variance of the precision matrix.}
\label{fig7}
\end{centering}
\end{figure}

The estimator of the precision matrix is distributed as the inverse-Wishart distribution \citep{Wishart1928,Press1982} and an analytical expression for the unbiased variance of the precision matrix has been derived in \citet{Taylor2013}:
\begin{equation}\label{variance_of_precision}
\sigma^2(\widehat{\mathcal{P}}_{ij})=A\left[(N_s-N_d)\mathcal{P}_{ij}^2+(N_s-N_d-2)\mathcal{P}_{ii}\mathcal{P}_{jj}\right],
\end{equation}
where $A=(N_s-N_d-1)^{-1}(N_s-N_d-4)^{-1}$. As in the case of the covariance errors we test this relation along the diagonal components and compute the ratio
\begin{equation}\label{err_prec_scaling}
\epsilon_{\mathcal{P}}=\sqrt{\frac{\sum_i\sigma^2(\widehat{\mathcal{P}}_{ii})}{\sum_i\langle\widehat{\mathcal{P}}_{ii}\rangle^2}}=\sqrt{\frac{2}{N_s-N_d-4}},
\end{equation}
in the range $0.11<k\,[h\,{\rm Mpc^{-1}}]<1.22$ which we plot in Fig.~\ref{fig6} as function of $N_s$. Differently from the covariance errors we notice that deviations from the expected scaling do not exceed $5\%$ level for $N_s>1200$. Similarly the off-diagonal components do not show significant departures above $5\%$ independently of the level of correlation as shown in Fig.~\ref{fig7}. 

\section{Covariance Estimation Errors and Parameter Forecast}\label{sec_fisher}
We perform a Fisher matrix analysis \citep{Tegmark1997} to investigate the impact of sample covariance errors on the cosmological parameter uncertainties. As already mentioned in Section~\ref{sec2} non-linearities induce deviations of the likelihood of the power spectrum from a Gaussian distribution \citep{Blot2015}, \citep[for the case of primordial non-Gaussianity see][]{Kalus2015}. Moreover, when inferring parameters from a likelihood constructed with an estimated covariance the likelihood function is not Gaussian even for Gaussian distributed data \citep{Sellentin2015}. Understanding the impact of these effects will require a full likelihood data analysis or an extension to the Fisher matrix formalism \citep{Sellentin2014} that are beyond the scope of this work. Here, we want to focus on the impact of the covariance errors on the cosmological parameter uncertainties, thus we limit for simplicity to a Gaussian approximation of the likelihood.

We consider a survey with the same volume of the simulations $(656.25\,h^{-1}\,\textrm{Mpc})^3$ and a mean galaxy density similar to the one expected by the Euclid surveys \citep{Laureijs2011}. Despite the fact that we use a much smaller volume than the Euclid surveys, this analysis allows us to understand what is the impact of non-linearities in a shot noise regime that is compatible with future galaxy surveys.
In the following we assume as fiducial cosmology the flat $\Lambda$CDM model best-fit to the WMAP-7 years used for the DEUS-PUR simulations (see Table~\ref{tab:cosmopar}). We consider measurements of the galaxy power spectrum at $N_z=5$ redshifts in the range $0.5\le z\le 2$ and mean galaxy number densities $\bar{n}_g(z)$ as given in Table~\ref{tab1}. At each redshift we consider $N_d=250$ band power measurements in the range $0.03<k\,[h\,{\rm Mpc}^{-1}]<1.22$.
\begin{table}
\centering
\begin{tabular}{c|c|c|c|c}
\hline
\ $h$ & $\Omega_m h^2$ & $\Omega_b h^2$ & $n_s$ & $\sigma_8$ \\
\hline
\ $0.72$ & $0.1334$ & $0.02258$ & $0.963$ & $0.801$ \\
\hline
\end{tabular}
\caption{DEUS-PUR cosmological model parameter values.}\label{tab:cosmopar}
\end{table}

\begin{table}
\begin{center}
\begin{tabular}{|c|c|}
\hline
$z$ & $\bar{n}_g(z)$ \\
\hline
$0.5$ & $4.2\times10^{-3}$ \\
$0.7$ & $2.99\times10^{-3}$ \\
$1.0$ & $1.81\times10^{-3}$ \\
$1.5$ & $0.77\times10^{-3}$ \\
$2.0$ & $0.15\times10^{-3}$ \\
\hline
\end{tabular}
\caption{Redshifts and mean galaxy number densities as expected from a Euclid-like survey \citep[][]{Laureijs2011}.}\label{tab1}
\end{center}
\end{table}

We model the galaxy power spectrum at a given redshift $z$ as: 
\begin{equation}
P_g(k;z)=b_z^2 P(k;z),\label{pgal} 
\end{equation}
where $b_z$ is a constant bias parameter, $P(k;z)$ is the non-linear matter power spectrum at redshift $z$ while we model the galaxy power spectrum covariance matrix as in \citet[][]{Takahashi2009}:
\begin{align}\label{eq:covg}
\textrm{cov}_g(k_i,k_j;z)&=b_z^4\,\widehat{\mathcal{C}}_{ij}(z)+2 b_z^2[P(k_i;z)P(k_j;z)]^{1/2}\bar{n}^{-1}_g(z) \nonumber\\
&+\bar{n}^{-2}_g(z), 
\end{align}
where we use the matter power spectrum covariance matrix at redshift $z$ computed from the DEUS-PUR simulations as $\widehat{\mathcal{C}}_{ij}(z)$ and the mean of the matter power spectra over the simulation set as $P(k;z)$. Here we have neglected the effect of Redshift Space Distortions (RSD), the study of which would require a modelling of the anisotropic power spectrum in the non-linear regime that is beyond the scope of this work \citep[see][ for a model of Gaussian covariance for the anisotropic galaxy power spectrum]{Grieb2015}.

We aim to forecast the impact of covariance errors on the following set of cosmological parameters: the cosmic matter density $\Omega_m$, the constant dark energy equation of state $w$, the normalisation of the power spectrum $\sigma_8$, the scalar spectral index $n_s$ and the cosmic baryon density $\Omega_b$. In addition we consider constant bias parameters $b_{z_l}$ with $l=1,N_z$ for each redshift with fiducial value set to $1$. We define the following vector of model parameters $\vec{\theta}=\{\Omega_m,w,\sigma_8,n_s,\Omega_b,b_{z_1},..,b_{z_{N_z}}\}$.

In the case of Gaussian distributed data the likelihood is given by Eq.~(\ref{likelihood}). Expanding the likelihood to second order around the fiducial model parameter values we can obtain a lower bound estimate of the expected model parameter errors from the Fisher matrix, that reads as:
\begin{equation}
F_{\alpha\beta}=\sum_{l=1}^{N_z} \sum_{i,j=1}^{N_d}\frac{\partial P_g}{\partial\theta_\alpha}(k_i;z_l)\frac{\partial P_g}{\partial\theta_\beta}(k_j;z_l)\textrm{cov}_g^{-1}(k_i,k_j;z_l),\label{fisher}
\end{equation}
where we have neglected correlations among different redshift bins. We compute the derivatives of the non-linear matter power spectrum using a five-point stencil approximation:
\begin{eqnarray}
\frac{\partial P_g}{\partial\theta_\alpha}&\approx&\frac{2}{3}\frac{P_g(\hat\theta_\alpha+\Delta\theta_\alpha)-P_g(\hat\theta_\alpha-\Delta\theta_\alpha)}{\Delta\theta_\alpha}+\nonumber\\
&+&\frac{P_g(\hat\theta_\alpha+2\Delta\theta_\alpha)-P_g(\hat\theta_\alpha-2\Delta\theta_\alpha)}{12\Delta\theta_\alpha},
\end{eqnarray}
where $\hat\theta_\alpha$ is the fiducial parameter value and $\Delta\theta_\alpha=0.05\,\hat\theta_\alpha$. To account for the effect of non-linearities in the finite derivatives we compute the non-linear $P(k;z)$ for the different cosmological parameter values using the emulator PkANN\footnote{http://zuserver2.star.ucl.ac.uk/fba/PkANN/} \citep{Agarwal2014}. The emulator reproduces the non-linear matter power spectra of N-body simulations to few percent accuracy over the range of scales and redshifts of interest for different combinations of the cosmological model parameters. Let us note here that the PkANN emulator does not allow to fix the Hubble parameter $h$ but, given the other cosmological parameters, its value is computed by fitting the combined WMAP 7-year and BAO constraints on the acoustic scale \citep[for more details see][]{Agarwal2014}.

\begin{figure}
\begin{centering}
\includegraphics[scale=0.5]{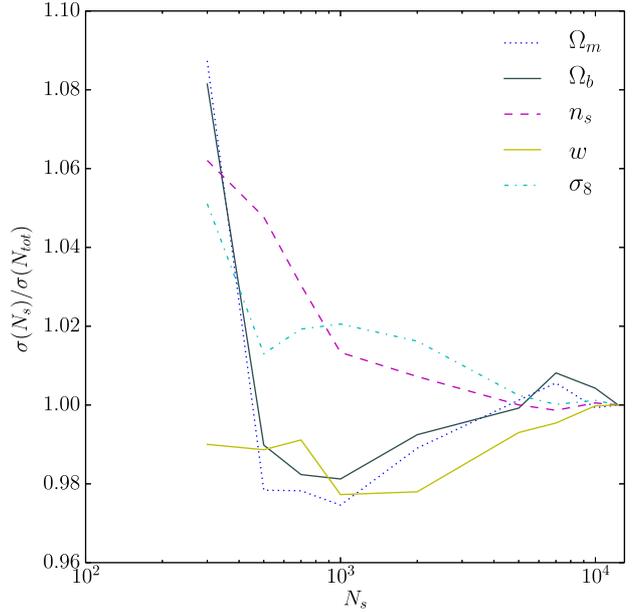}
\caption{Fractional error contribution to the cosmological parameter errors of the numerically estimated covariance as function of the number of simulations relative to the errors obtained using the covariance from the full DEUS-PUR sample.}
\label{fig10}
\end{centering}
\end{figure}

In Fig.~\ref{fig10} we plot the ratio of the $1\sigma$ marginalised model parameter errors $\sigma_{\vec{\theta}}$ as function of the number of simulations with respect to the errors obtained using the full DEUS-PUR sample. We can see that the fractional error contribution of the sample covariance to the parameter errors reduces to sub-percent level for $N_s>5000$. Note that we have assumed here that the covariance estimated with 12288 simulations is the true covariance, so to validate this requirement one would need $\gg 10^4$ simulations.

\subsection{On the linear approximation for the covariance}
\begin{figure*}
\begin{minipage}{17cm}
\includegraphics[scale=0.47]{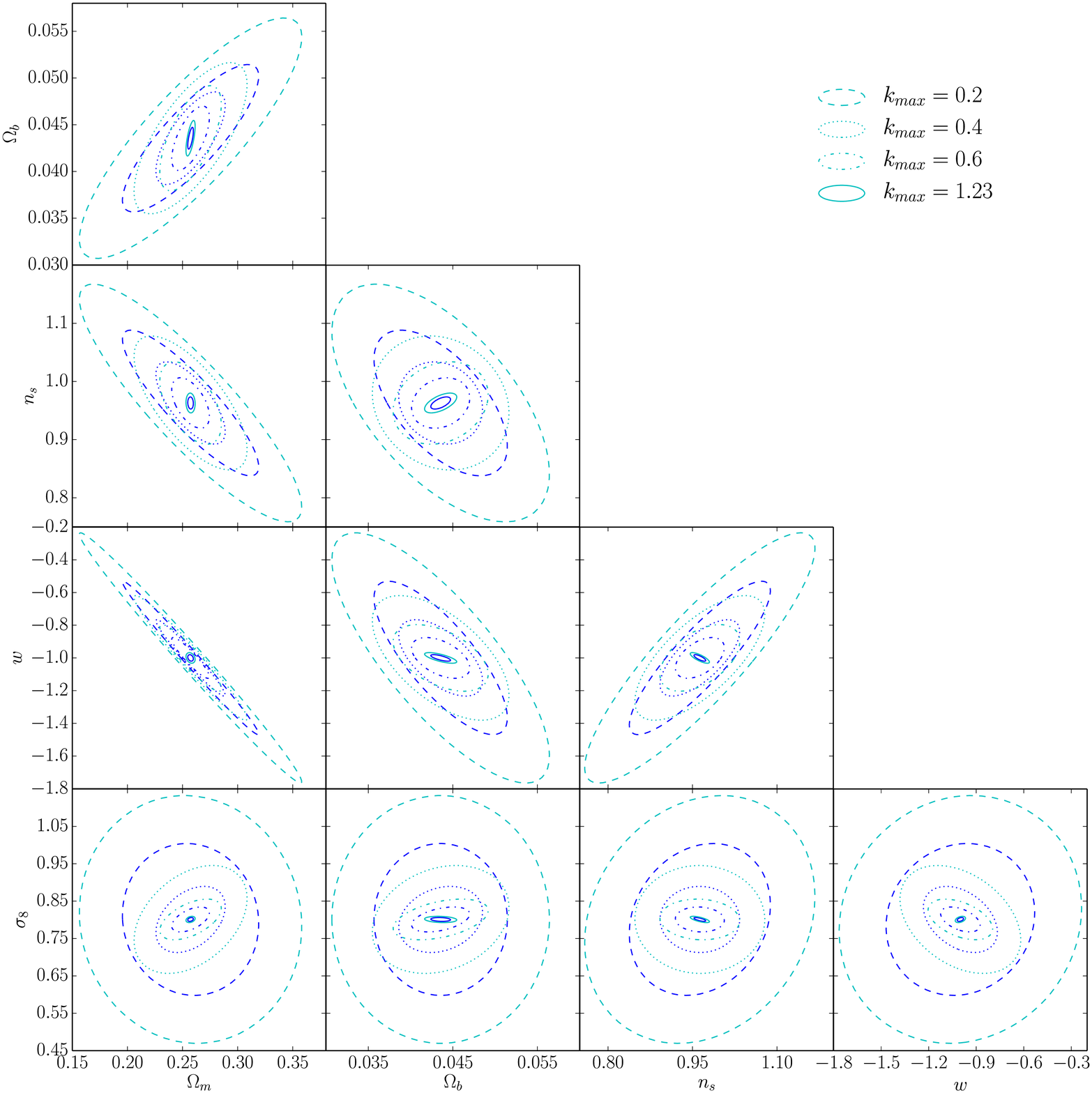}
\caption{\label{euclid_lin}Fisher matrix $2$-parameter $1$ and $2\sigma$ contours with $k_{max}=0.2$ $h\,{\rm Mpc}^{-1}$ (dashed lines), $k_{max}=0.4$ $h\,{\rm Mpc}^{-1}$ (dotted lines), $k_{max}=0.6$ $h\,{\rm Mpc}^{-1}$ (dash-dotted lines) and $k_{max}=1.22$ $h\,{\rm Mpc}^{-1}$ (solid lines) for various combination of the cosmological parameters marginalised over the constant redshift bin bias parameters.}
\end{minipage}
\end{figure*}

In the literature of Fisher forecasting it is common to consider the power spectrum only on the largest scales, e.g. $k<0.2$ $h\,{\rm Mpc}^{-1}$, where it is believed to be well approximated by the linear power spectrum. This is because accurate modelling of non-linear effects is necessary to push the analysis to smaller scales and the flattening of the signal-to-noise ratio at large $k$ values hints to a reduced information content in the power spectrum at these scales \citep{Angulo2008,Smith2009,Takahashi2009,Blot2015}. 
To show what is the impact of considering non-linear scales we compare in Fig.~\ref{euclid_lin} the marginalised 2-parameter $1$ and $2\sigma$ contours for $k_{max}=0.2$ $h\,{\rm Mpc}^{-1}$ (dashed lines), $k_{max}=0.4$ $h\,{\rm Mpc}^{-1}$ (dotted lines), $k_{max}=0.6$ $h\,{\rm Mpc}^{-1}$ (dash-dotted lines) and $k_{max}=1.22$ $h\,{\rm Mpc}^{-1}$ (solid lines). These contours are not to be interpreted as forecasts, since as we stated above the volume that we consider is much smaller than the one expected from future galaxy surveys, nevertheless these show that pushing the analysis to non-linear scales largely improves constraints on cosmological parameters. The differences can be inferred more clearly in Table~\ref{tab3} where we report the marginalised $1\sigma$ errors on the cosmological parameters for different values of $k_{max}$.

\begin{figure*}
\begin{minipage}{17cm}
\begin{center}
\includegraphics[scale=0.47]{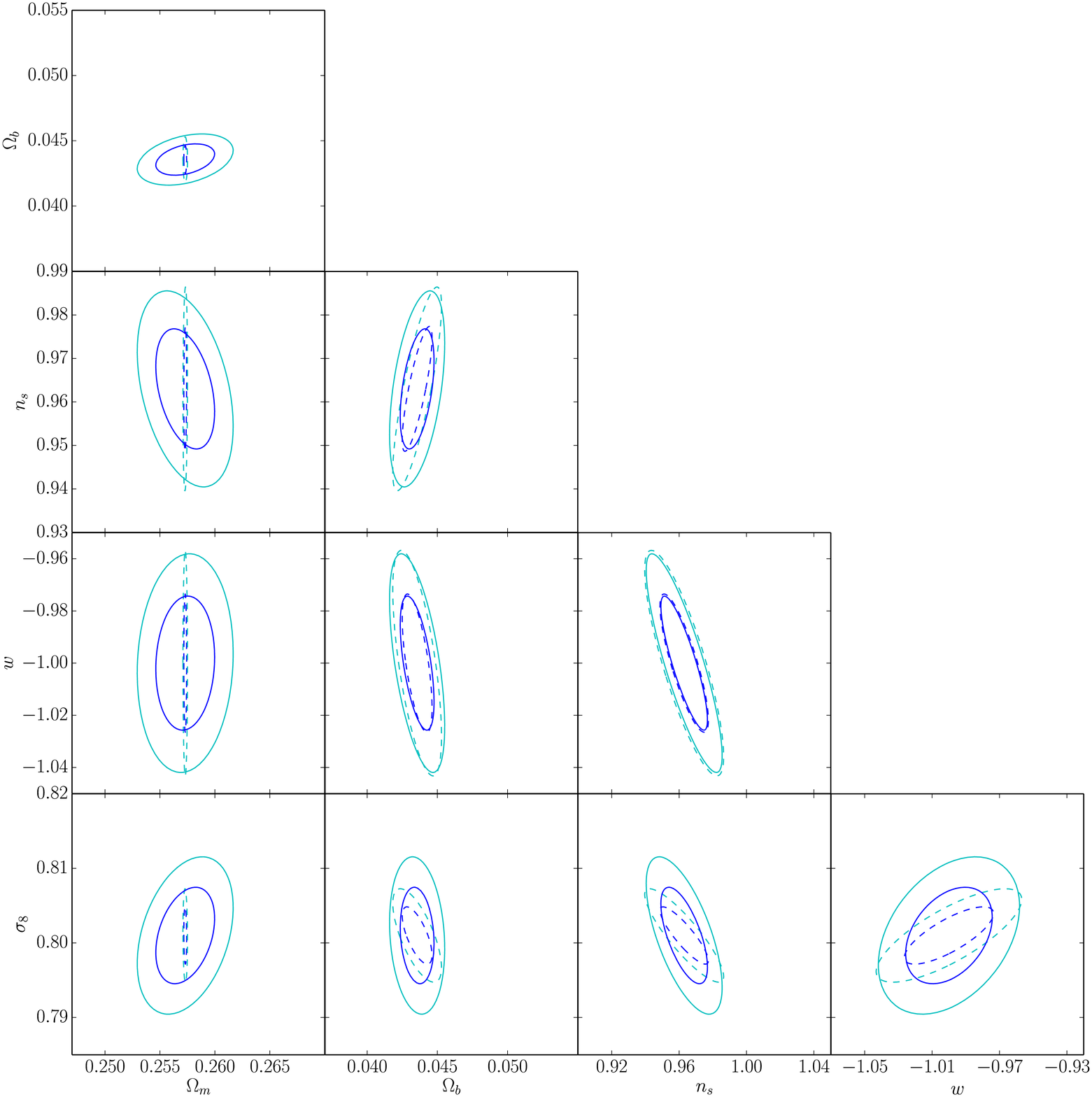}
\caption{\label{euclid}Fisher matrix $2$-parameter $1$ and $2\sigma$ contours in the case of a Gaussian linear matter power spectrum covariance (dashed lines) and of the fully non-linear covariance from the DEUS-PUR simulation ensemble (solid lines) for various combination of the cosmological parameters marginalised over the constant redshift bin bias parameters.}
\end{center}
\end{minipage}
\end{figure*}

In the linear regime the matter density field is Gaussian and the power spectrum estimator at a given $k$ is $\chi^2$-distributed with $N_k$ degrees of freedom, where $N_k\approx k^2\Delta k \,V/(2\pi^2)$ is the number of modes in a bin of width $\Delta k$. Moreover, each Fourier mode evolves independently so that the matter power spectrum covariance is diagonal and reads as:
\begin{equation}\label{covdev}
\mathcal{C}_{ij}=\frac{2}{N_{k_i}}P_{lin}^2(k_i)\delta_{ij},
\end{equation}
where the diagonal components are proportional to the linear matter power spectrum of the fiducial cosmology $P_{lin}$ \citep[see e.g.][]{Jeong2009}. Using this approximation in non-linear regime, as it is often done, leads to an underestimation of the errors on the cosmological parameters. To show this, we perform a Fisher analysis by substituting the matter power spectrum covariance $\widehat{\mathcal{C}}_{ij}$ with Eq.\eqref{covdev} and we compare the results to those obtained with the covariance from the full ensemble of DEUS-PUR simulations.
In Fig.~\ref{euclid} we plot the marginalised 2-parameter $1$ and $2\sigma$ contours obtained with the linear approximation Eq.\eqref{covdev} (dashed lines) and with the fully non-linear covariance from the DEUS-PUR simulations (solid lines). The differences can be inferred more clearly in Table~\ref{tab2} where we report the marginalised $1\sigma$ errors on the cosmological parameters for the two cases. The discrepancy is maximal in the case of $\Omega_m$, for which the use of a linear covariance underestimates the $1\sigma$ error by a factor of $\sim 17.7$. Smaller differences occur for $\sigma_8$ and $\Omega_b$ though still exceeding the $\sim 20\%$ level.

\begin{table}
\begin{center}
\begin{tabular}{|c|c|c|c|c|}
\hline
$k_{max}$ & $0.2$ & $0.4$ & $0.6$ & $1.22$\\
\hline
$\Omega_m$ & $4.1\times 10^{-2}$ & $2.1\times 10^{-2}$ & $1.1\times 10^{-2}$ & $1.6\times 10^{-3}$\\
$\Omega_b$ & $5.2\times 10^{-3}$ & $3.3\times 10^{-3}$ & $2.3\times 10^{-3}$ & $7.7\times 10^{-4}$\\
$n_s$ & $8.2\times 10^{-2}$ & $4.6\times 10^{-2}$ & $2.9\times 10^{-2}$ & $6.9\times 10^{-3}$\\
$w$ & $3.1\times 10^{-1}$ & $1.5\times 10^{-1}$ & $8.2\times 10^{-2}$ & $1.3\times 10^{-2}$\\
$\sigma_8$ & $1.3\times 10^{-1}$ & $5.8\times 10^{-2}$ & $2.2\times 10^{-2}$ & $3.4\times 10^{-3}$\\
\hline
\end{tabular}
\caption{Marginalised $1\sigma$ Fisher matrix errors on the cosmological parameters obtained using the DEUS-PUR covariance for different values of $k_{max}$ [$h\,{\rm Mpc}^{-1}$].}\label{tab3}
\end{center}
\end{table}

\begin{table}
\begin{center}
\begin{tabular}{|c|c|c|}
\hline
$\theta$ & $\sigma^{G}_\theta$ & $\sigma^{NG}_\theta$\\
\hline
$\Omega_m$ & $9.6\times 10^{-5}$ & $1.7\times 10^{-3}$\\
$\Omega_b$ & $6.1\times 10^{-4}$ & $7.7\times 10^{-4}$\\
$n_s$ & $7.5\times 10^{-3}$ & $6.9\times 10^{-3}$\\
$w$ & $1.4\times 10^{-2}$ & $1.3\times 10^{-2}$\\
$\sigma_8$ & $2.3\times 10^{-3}$ & $3.4\times 10^{-3}$\\
\hline
\end{tabular}
\caption{Marginalised $1\sigma$ Fisher matrix errors on the cosmological parameters obtained using a linear covariance (second column) and the DEUS-PUR covariance (third column).}\label{tab2}
\end{center}
\end{table}

\section{Conclusion}\label{conclu}
Future measurements of the matter power spectrum will require accurate estimates of the covariance matrix. Independent realisations of the data can be used to estimate the covariance. However, sampling errors will propagate in the cosmological parameter inference, thus potentially limiting the expected performance of future galaxy survey measurements. 
In this work we have studied the impact of non-Gaussian errors on the sample covariance due to the non-linearities of the matter density field during the late-time regime of clustering. 
To this purpose we have used a large ensemble of N-body simulations. We find that covariance sampling errors increasingly deviate from Gaussian expectations above 10\% level on scales $k\gtrsim 0.3\, h\,{\rm Mpc}^{-1}$. To assess the impact on the cosmological parameter uncertainties from future surveys such as Euclid we have performed a Fisher analysis forecast. We have shown that including non-linear scales largely improves the constraints on cosmological parameters. We have also shown that using a linear matter power spectrum covariance at these scales will significantly underestimate errors, which emphasises the need of using a fully non-linear covariance to correctly analyse future data. However, our analysis also indicates that to reduce the effect of sample covariance errors on parameter uncertainties at sub-percent level a very large ensemble of numerical simulations will be needed. Advancements in numerical N-body simulations techniques may render this task tractable in the future. Nevertheless the development of theoretical approaches capturing the relevant features of the non-linear collapse of matter on the scale of interest should be pursued to facilitate the statistical analysis of future data as well as to infer information on the late-time clustering of matter.

\section*{Acknowledgements}
We are grateful to Will Percival and Cong Ma for useful discussions. The research leading to these results has received funding from the European Research Council under the European Union's Seventh Framework Programme (FP/2007-2013) / ERC Grant Agreement n. 279954. L.A. acknowledges financial support from DFG through the project TRR33 ``The Dark Universe". L.B. acknowledges support from the Spanish Ministerio de Economia y Competitividad grant ESP2014-58384-C3-1-P.

\end{document}